\documentclass[twocolumn,preprintnumbers,amsmath,amssymb]{revtex4}

\usepackage{graphicx}
\usepackage{dcolumn}
\usepackage{bm}
\usepackage{subfigure}
\usepackage{color}
\usepackage{mathrsfs}
\usepackage{ascmac}
\newcommand{\Cr}{CrNb$_3$S$_6$}
\newcommand{\Yb}{YbNi$_3$Al$_9$} 

\begin{document}

\title{Finite-Temperature Properties of Three-Dimensional Chiral Helimagnets}

\author{Misako Shinozaki$^1$}
\email{shinozaki@vortex.c.u-tokyo.ac.jp}
\author{Shintaro Hoshino$^1$}
\author{Yusuke Masaki$^2$}
\author{Jun-ichiro Kishine$^3$}
\author{Yusuke Kato$^1$}
\affiliation{$^1$Department of Basic Science, The University of Tokyo, Meguro-ku, Tokyo 153-8902, Japan}
\affiliation{$^2$Department of Physics, The University of Tokyo, Bunkyo-ku, Tokyo 113-0033, Japan}
\affiliation{$^3$The Open University of Japan, Mihama-ku, Chiba 261-8586, Japan}
\date{\today}

\begin{abstract}
We study a three-dimensional (3d) classical chiral helimagnet at finite temperatures through analysis of a spin Hamiltonian, which is defined on a simple cubic lattice and consists of Heisenberg exchange, mono-axial Dzyaloshinskii-Moriya interactions and Zeeman energy due to magnetic field applied in the plane perpendicular to the helical axis. 
We take account of quasi-two-dimensionality of a known mono-axial chiral helimagnet \Cr\ and we adopt three methods: (i) a conventional mean-field (MF) analysis which we call 3dMF method, (ii) a hybrid method called 2dMC-1dMF, which is composed of a classical Monte Carlo (MC) simulation and MF approximation applied respectively to the intra- and inter-plane interactions, and (iii) a simple-MC simulation (3dMC) at zero field. 
Temperature dependence of magnetization calculated by the 3dMF method shows a cusp-like structure similar to that observed in experiments. 
In the absence of magnetic field, both 2dMC-1dMF and 3dMC yield close values of transition temperature. 
The 2dMC-1dMF provides a quantitative description of thermodynamic properties even under external field at an accessible numerical cost.
\end{abstract}
\maketitle

\section{\label{sec:level1}Introduction}

Materials with chirality, i.e., left- or right-handedness, have been attracting continued attention in condensed matter physics. Such systems have been found in magnetic compounds, where the chirality in crystal structure causes an antisymmetric exchange interaction called Dzyaloshinskii-Moriya (DM) interaction~\cite{Dzyaloshinskii_D,Moriya_M}. The competition between the symmetric Heisenberg exchange and the mono-axial DM interactions gives rise to a long-period helical magnetic texture~\cite{Dzyaloshinskii, Dzyaloshinskii2, Dzyaloshinskii3,Izyumov85} called chiral helimagnetic state~\cite{Kishine_S_T}. Under external magnetic field perpendicular to the helical axis, chiral helimagnet turns into a periodic array of magnetic soliton kinks, called chiral soliton lattice (CSL)~\cite{Kishine_S_T}. The CSL is intrinsic to chiral magnet in the sense that it does not appear in purely Yoshimori-type (i.e., non-chiral) helimagnets~\cite{Yoshimori}.
Dzyaloshinskii~\cite{Dzyaloshinskii} developed a Ginzburg-Landau theory for CSL and elucidated the critical behavior near the phase boundary between CSL phase and paramagnetic (=forced ferromagnetic) region~\cite{Dzyaloshinskii3}. 

Moriya and Miyadai~\cite{Moriya}, and Miyadai et al.~\cite{Miyadai} argued the existence of helimagnetic and CSL structure in \Cr, respectively, on the basis of neutron scattering experiments and magnetization curve. In this compound, magnetic moments are associated with the intercalated Cr$^{3+}$ ions between the NbS$_2$-layers, each of which have magnetic moment 2.9$\mu_{\rm B}$, which is nearly full moment for spin 3/2 electrons. 
Recently, Togawa et al. directly observed the CSL structure in \Cr\ using Lorentz transmission electron microscopy, and found that the chiral magnetic structures are very stable against crystal defects~\cite{Togawa}. With this result, \Cr\ is recognized as one of the candidates of new magnetic devices~\cite{Togawa2}. 
In addition to \Cr, \Yb\ was also discovered as another mono-axial chiral helimagnet with heavy-electron behavior, where the periodicity of the helix is comparable to lattice constant~\cite{Ohara2011, Miyazaki, Ohara}.

In recent theoretical studies~\cite{Kishine_S_T,Kishine_MTcusp, Kishine_Hc_perp, Kishine_Ovchinnikov, Kishine_lattice}, magnetic properties of~\Cr\ have been studied with a one-dimensional (1d) continuum model, which is called chiral sine-Gordon model. The analysis of this model has revealed characteristic magnetization curves, which have been observed in experiments on \Cr~\cite{Moriya,Miyadai,Ghimire,Chapman,Dyadkin} and \Yb~\cite{Miyazaki,Ohara,Ohara2011}. 
Temperature effect of the spin moment has been included phenomenologically \cite{Dzyaloshinskii3,Kishine_MTcusp} and magnetization, entropy and specific heat have been discussed.

While these studies using chiral sine-Gordon model are in good agreement with experiments, there are still remaining issues to be addressed as described below. 
Miyadai et al. argued that, in \Cr, spins in an $ab$-plane are strongly coupled, so that they can be treated as a single spin with large magnetic moment~\cite{Miyadai}. 
This was one of the motivations of analyzing 1d systems, so that the analysis using the chiral sine-Gordon model actively was conducted recently. 
The existing theory is, however, valid only at zero temperature for arbitrary fields or in the vicinity of the phase boundary between paramagnetic and CSL at finite temperatures. 
The theory applicable in the wider temperature region is highly required to describe overall thermodynamic properties that include crossover phenomena as well as the phase transition.

Another issue is on the origin of the transition temperature of \Cr\ that amounts to $T_{\rm c}(H=0)$=127~K~\cite{Miyadai}. 
This rather high-$T_{\rm c}$ has attracted much attention from the point of view of the exchange interaction, so that a quantitative estimation of the energy scales in this crystal is required.

To address these issues, we consider chiral helimagnetic systems through classical spin model defined on a three-dimensional (3d) lattice using three methods. 
As a simple and prerequisite analysis, we first present the result of the conventional mean-field (MF) method, which we call 3dMF method. 

Second, we take account of the spin-correlation effect in the two-dimensional (2d) layer perpendicular to the helical axis with a Monte Carlo (MC) method and treat interlayer coupling via the MF theory. We call this method 2dMC-1dMF method.  By this method, we aim at quantitative description of finite-temperature properties of quasi-2d systems such as \Cr. The idea of this method is based on an earlier work~\cite{Scalapino}, where the generalized Ginzburg-Landau theory was constructed to describe the phase transition in quasi-1d Ising and Heisenberg models. 

Third, we perform the MC simulation for the 3d systems, which is the most proper method for the quantitative analysis among these three methods. We call it 3dMC method in this paper and apply it to the case without magnetic field.

Finally, we estimate the interaction parameters in \Cr\ by comparison between the experimental data and our numerical results of 2dMC-1dMF and 3dMC. 

In the next section, we introduce a 3d Heisenberg model with the DM interaction on a simple cubic lattice, and formulate  the MF theory. 
In Sec.~\ref{section:numerical methods}, we describe the simulation details of the MF and MC methods, and show the numerical results in Sec.~\ref{section:results}. 
In Sec.~\ref{section:discussion}, the relevance of our results to \Cr\ will be discussed, and we summarize the paper in Sec.~\ref{section:conclusion}.

\section{Formulation\label{section:formulation}}
\subsection{Model}
We consider the following Hamiltonian for the chiral helimagnet on a 3d cubic crystal:
\begin{align}
	{\cal H} = -J^{\bot}\sum_{\bm{i}} {\bm S}_{\bm{i}}\cdot ({\bm S}_{\bm{i}+\hat{\bm x}} + {\bm S}_{\bm{i}+\hat{\bm y}})-J^{\parallel}\sum_{\bm{i}} {\bm S}_{\bm{i}} \cdot {\bm S}_{\bm{i}+\hat{\bm z}}
				\notag\\
		- D \sum_{\bm{i}} ({\bm S}_{\bm{i}}\times{\bm S}_{\bm{i}+\hat{\bm z}})\cdot \hat{\bm z}
		- H^\bot \sum_{\bm{i}}  \bm S_{\bm{i}} \cdot \hat {\bm x}
		.\label{eq:hamiltonian}
\end{align}
Each site on the cubic lattice is specified by a dimensionless vector $\bm{i}=i_x\hat{\bm{x}}+i_y \hat{\bm{y}}+i_\parallel \hat{\bm{z}}$ with integers $i_x$,~$i_y$,~$i_\parallel$. The basis vectors $\hat{\bm{x}}$, $\hat{\bm{y}}$ and $\hat{\bm{z}}$ denote, respectively, the unit vectors of $x$,~$y$,~$z$ directions. We regard the cubic lattice as a set of layers, each of which is labeled by $i_\parallel$. 
The symbol ${\bm S}_{\bm{i}}=(S^x_{\bm{i}},S^y_{\bm{i}},S^z_{\bm{i}})$ represents the classical Heisenberg spin with the magnitude $S$ at site $\bm{i}$. 
We denote by $J^\bot>0$ a ferromagnetic exchange interaction between neighboring spins within the same layer, and $J^\parallel>0$ and $D$, respectively, are ferromagnetic exchange interaction and magnitude of DM interaction between a pair of neighboring spins in adjacent layers. The superscripts $\parallel$ and $\bot$ represent parallel and perpendicular to the $z$-axis, respectively. 

In earlier theoretical studies using the 1d chiral sine-Gordon model, intra-layer exchange interaction $J^\bot>0$ is assumed to be much larger than other energy scales such as $J^\parallel$ and $D$, so that all spins can be regarded as fully polarized within each layer. In this case, $\bm{S}_{\bm{i}}$ depends only on $i_\parallel$ and can be rewritten as $\bm{S}_{i_\parallel}$. The Hamiltonian [Eq.~\eqref{eq:hamiltonian}] is then reduced to 
\begin{align}
	{\cal H}/N_{\rm 2d} = & - N_z J^{\bot} S^2 -J^{\parallel}\sum_{i_\parallel} {\bm S}_{i_\parallel} \cdot {\bm S}_{i_\parallel +1}
				\notag\\
		&- D \sum_{i_\parallel} ({\bm S}_{i_\parallel}\times{\bm S}_{i_\parallel+1})\cdot \hat{\bm z}
		- H^\bot \sum_{i_\parallel}  \bm S_{i_\parallel} \cdot \hat {\bm x}
		\label{eq:1d-hamiltonian}
\end{align}
at zero temperature, where $N_{\rm 2d}$ is the number of sites in a 2d layer and $N_z$ is the number of layers. 
In the absence of the magnetic field, the combination between $J^\parallel$ and $D$ in the second and third terms in Eq.~\eqref{eq:1d-hamiltonian} generates the chiral helimagnetic structure, which is characterized by the uniform invariant
\begin{align}
	S^x_{i_{\parallel}+1}S^y_{i_\parallel}-S^y_{i_{\parallel}+1}S^x_{i_\parallel},
	\label{eq:pitch}
\end{align}
which is {\it independent} of $i_\parallel$. 
The magnetic field $H^\bot \hat{\bm{x}}$ induces the CSL structure, in which the value of [Eq.~\eqref{eq:pitch}] modulates periodically as a function of $i_\parallel$.

\subsection{3dMF formulation\label{subsection:3d_mean_field}}

We investigate the finite-temperature properties on the basis of the 3d Hamiltonian [Eq.~\eqref{eq:hamiltonian}]. 
Let us apply the MF theory to the 3d system. To this end, we write the site index as $\bm{i}=\bm{i}_\perp+i_\parallel\hat{\bm{z}}$ where $\bm{i}_\perp=i_x \hat{\bm{x}}+i_y \hat{\bm{y}}$. The system is then described by the following single-site Hamiltonian:
\begin{align}
	{\cal H}_{i_\parallel, {\bm i}_\perp}^{\rm MF} 
	= - {\bm H}^{\rm eff}_{i_\parallel}
		\cdot \bm S_{i_\parallel,{\bm i}_\perp}
	+ C_{i_\parallel},
	\label{eq:3d_MF}
\end{align}
where the effective field is given by
\begin{align}
	{\bm H}^{\rm eff}_{i_\parallel} = 
	& J^\parallel (\bm M_{i_{\parallel}+1} + \bm M_{i_{\parallel}-1}) + 4J^{\bot} \bm M_{i_\parallel}
		\notag\\
		&+ D (\bm M_{i_{\parallel}+1} - \bm M_{i_{\parallel}-1}) \times \hat {\bm z}
		+H^\perp \hat {\bm x},
	\label{eq:mf_heff}
\end{align}
and the constant term is defined by 
\begin{align}
	C_{i_\parallel} = \frac{{\bm H}^{\rm eff}_{i_\parallel} - H^\bot\hat{\bm x}}{2} \cdot {\bm M}_{i_\parallel}.
	\label{C2}
\end{align} 
We have defined the thermal average of the spin moment $\bm M_{i_\parallel} = \langle \bm S_{i_\parallel, \bm{i}_\perp} \rangle$  for 2d Heisenberg model, which is explicitly written as
\begin{align}
	\frac{{\bm M}_{i_\parallel}}{S} =
	\left[
		\coth(\beta S|{\bm H}^{\rm eff}_{i_\parallel}|) -\frac{1}{\beta S|{\bm H}^{\rm eff}_{i_\parallel}|}
	\right]
	\frac{{\bm H}^{\rm eff}_{i_\parallel}}{|{\bm H}^{\rm eff}_{i_\parallel}|},
	\label{eq:mf_M}
\end{align}
where $\beta=1/T$ with $k_{\rm B}=1$. Here, ${\bm H}^{\rm eff}_{i_\parallel}$ and ${\bm M}_{i_\parallel}$ are independent of the index $i_\bot$. The free energy of the MF Hamiltonian is given by 
\begin{align}
	\frac{F^{\rm 3dMF} }{N_{\rm 2d}}
	= -\frac{1}{\beta} \sum_{i_\parallel} \log 
		\left(
			\frac{\sinh(\beta S|{\bm H}^{\rm eff}_{i_\parallel}| )}{\beta S |{\bm H}^{\rm eff}_{i_\parallel}|}
		\right)
	+\sum_{i_\parallel} C_{i_\parallel}.
	\label{eq:F_3dMF}
\end{align}

At zero field ($H^\perp = 0$), we assume that the helical magnetic structure is realized. 
In this case, $\bm H^{\rm eff}_{i_\parallel}$ is given in the form of helical field: 
\begin{align}
	&\bm H^{\rm eff}_{i_\parallel}=H^{\rm eff}\left(\cos(\theta i_\parallel +\Theta),\sin(\theta i_\parallel +\Theta),0\right)		\label{eq:Heff-1},
	\\
	&	
	H^{\rm eff} = 2M_{\rm h}(J^\parallel  \cos\theta + D \sin\theta), 
	\label{eq:Heff-2}
\end{align}
where $\theta$ is the angle between the spins located at the nearest neighbor layers and $\Theta$ is an overall phase factor. Here, $M_{\rm h}$ is the magnitude of the spin moment for the helical magnetic structure. For a given $M_{\rm h}$, the magnitude of the effective field, $H^{\rm eff}$, becomes maximum and the transition temperature is expected to be highest when $\theta=\alpha$ with
\begin{align}
	\alpha = \arctan(D/J^\parallel ).
\end{align}
When we apply the small test helical magnetic field 
\begin{align}
	\bm{H}_\alpha=H_\alpha\left(\cos(\alpha i_\parallel),\sin(\alpha i_\parallel),0\right),	\label{eq:Htheta}
\end{align}
we obtain $\Theta=0$ in Eq.~\eqref{eq:Heff-1} and 
\begin{align}
	M_{\rm h}= \chi^{\rm 2dMF} (H_\alpha + H^{\rm eff}),
	\label{eq:2}
\end{align}
where $\chi^{\rm 2dMF}$ is the uniform susceptibility of the 2d Heisenberg model without fields given by $\chi^{\rm 2dMF} = S^2/(3T-4J^\bot S^2)$.
Substituting Eq.~\eqref{eq:Heff-2} with $\theta=\alpha$ into Eq.~\eqref{eq:2}, the helical susceptibility $\chi^{\rm 3dMF}_{\rm{h}} \equiv {\partial M_{\rm h}}/{\partial H_\alpha}$ for the 3d system is obtained by
\begin{align}
	\chi^{\rm 3dMF}_{\rm{h}} 
		 &=\frac{\chi^{\rm 2dMF}} { 1 - 2\chi^{\rm 2dMF}\sqrt{J^{\parallel 2} + D^2}}. 
	\label{eq:chi_3d}
\end{align}
The transition temperature $T_{\rm c} (H^\bot=0)$ is determined from the condition $\chi^{\rm 3d}_{\rm{h}} \to \infty$.
Then, we can derive the analytical expression of the transition temperature at zero field within the 3dMF approximation
\begin{align}
	T_{\rm c}^{\rm 3dMF}(H^\bot=0) =  \frac{4J^\bot + 2\sqrt{J^{\parallel 2}+D^2}}{3} S^2.
	\label{MF_Tc}
\end{align}

\subsection{2dMC-1dMF formulation}

Here, we apply the MF approximation only to the $z$-axis. The MF Hamiltonian at the $i_\parallel$-th layer is written as
\begin{align}
	{\cal H}_{i_\parallel}^{\rm MF} &= -J^{\bot}\sum_{\bm{i}_{\perp}} {\bm S}_{i_{\parallel}, \bm{i}_{\perp}} \cdot 
	({\bm S}_{i_{\parallel}, {\bm{i}_{\perp}}+\hat{\bm x}} + {\bm S}_{i_{\parallel}, {\bm{i}_{\perp}}+\hat{\bm y}}) 
	\nonumber \\
	&\ \ 
	- \bm H^{\rm eff}_{i_\parallel} \cdot \sum_{\bm{i}_\perp} \bm S_{i_{\parallel}, \bm{i}_{\perp}}
	+ C _{i_\parallel}
	\label{eq:part_MF}
\end{align}
with the following effective field: 
\begin{align}
	\bm H^{\rm eff}_{i_\parallel} 
	=& J^\parallel (\bm M_{i_{\parallel}+1} + \bm M_{i_{\parallel}-1}) 
		\notag\\
		&+ D (\bm M_{i_{\parallel}+1} - \bm M_{i_{\parallel}-1}) \times \hat {\bm z}
		+H^\perp \hat {\bm x},
	\label{eq:part_heff}
\end{align}
which acts uniformly on the $xy$-plane. The constant term is defined by the same form as Eq.~\eqref{C2}
\begin{align}
	C_{i_\parallel} = \frac{\bm H^{\rm eff}_{i_\parallel} - H^\perp \hat {\bm x}}{2}\cdot \bm M_{i_\parallel},
	\label{C}
\end{align}
with effective field defined by Eq.~\eqref{eq:part_heff}. Thus the problem is mapped onto a 2d system with uniform magnetic field. The free energy is calculated by
\begin{align}
	F^{\rm 1d MF} = N_zF^{\rm 2d}(0) + \sum_{i_\parallel}\Delta F_{i_\parallel} + \sum_{i_\parallel} C_{i_\parallel},
	\label{eq:F_1dMF}
\end{align}
where
\begin{align}
	\Delta F_{i_\parallel} = \int_0^{|{\bm H}_{i_\parallel}^{\rm eff}|} M^{\rm 2d}(H) dH
		= F^{\rm 2d}(|{\bm H}_{i_\parallel}^{\rm eff}|) - F^{\rm 2d}(0).
	\label{eq:DeltaF}
\end{align}
Here $M^{\rm 2d}(H)$ and $F^{\rm 2d}(H)$ are, respectively, the magnetization and the free energy of the 2d Heisenberg model with magnetic field $H$.

In the absence of the external field, the helical susceptibility $\chi^{\rm 3d}_{\rm{h}}$ is given by the same form as Eq.~\eqref{eq:chi_3d} using $\chi^{\rm 2d}$ instead of $\chi^{\rm 2dMF}$. Hence, the transition temperature at zero field can be obtained by solving the following equation:
\begin{align}
	 1-2\chi^{\rm 2d} \sqrt{J^{\parallel 2} + D^2} =0.
\label{eq:2d_MF}
\end{align}
We calculate the susceptibility of 2d Heisenberg Hamiltonian $\chi^{\rm 2d}$ using the MC simulation. This is a standard expression for transition temperature in quasi-low-dimensional systems~\cite{Scalapino}. Validity of this type of expression has been numerically confirmed for quasi-1d and quasi-2d antiferromagnetic quantum/classical Heisenberg models in Ref.~\cite{Yasuda}.

\section{Numerical Methods\label{section:numerical methods}}
\subsection{3dMF and 2dMC-1dMF methods}

We can obtain physical quantities at arbitrary temperatures by solving the self-consistent MF equations at each layer based on Eqs.~(\ref{eq:mf_heff})-(\ref{eq:mf_M}) or Eqs.~(\ref{eq:part_heff}) and (\ref{C}) numerically. In order to solve the mean-field equations efficiently, we prepare many initial conditions in the form
\begin{align}
	{\bm M}_{i_\parallel,{\bm i}_\bot} = S(\cos k{i_\parallel}, \sin k{i_\parallel}, 0), \ \   k = 2\pi w/N_z
	\label{eq:initial}
\end{align}
for $w=0,1,2,\cdots ,[w_0+1]$. Here $w_0 = N_z\alpha /(2\pi)$ is the winding number at zero field. The symbol $[x]$ denotes the greatest integer that is less than or equal to $x$. 
For each initial state, we solve the MF equations by calculating ${\bm H}^{\rm eff}_{i_\parallel}$ and ${\bm M}_{i_\parallel}$ iteratively. 
In the 2dMC-1dMF method, we have $|{\bm M}_{i_\parallel}|$ as a function of the magnetic field in advance using MC method, which is the same as $M^{\rm 2d}$ defined in the previous section.
After $10^5$ iterations, we pick up one of the final states that gives the lowest free energy, and calculate physical quantities with this state. The free energy can be calculated using Eq.~\eqref{eq:F_3dMF} for 3dMF method or Eq.~\eqref{eq:F_1dMF} for 2dMC-1dMF method. In the 2dMC-1dMF calculation, $\Delta F_{i_\parallel}$ is calculated in advance using the MC method. Here we note that $F^{\rm 2d}(0)$ in Eq.~\eqref{eq:DeltaF} is constant for fixed temperature; thus we do not have to consider this term for our purpose. 

After $10^5$ iterations, all relative errors between free energies before and after an iteration are less than $10^{-12}$ regardless of the initial condition. 
To evaluate the numerical precision, we calculate the following value in the 3dMF method:
\begin{align}
	\max_{i_\parallel,\ \mu=x,y,z} \left\{
		\left| 
			\frac{M_{i_\parallel}^\mu}{S}
			-
			\left[
				\coth(\beta S|{\bm H}^{\rm eff}_{i_\parallel}|) -\frac{1}{\beta S|{\bm H}^{\rm eff}_{i_\parallel}|}
			\right]
			\frac{(H^{\rm eff}_{i_\parallel})^\mu}{|{\bm H}^{\rm eff}_{i_\parallel}|}
		\right|
		\right\}
\end{align}
using the state $\{{\bm M}_{i_\parallel} \} $ which has the lowest free energy for a given set of parameters. We confirm that these values are less than $10^{-14}$.

\subsection{MC methods}

We perform the classical MC simulation to analyze the 3d chiral helimagnetic systems and to treat the 2d Heisenberg model which is used in the 2dMC-1dMF method. We employ the heat bath method~\cite{Miyatake,Loison} and the exchange MC method~\cite{Hukushima} in order to improve the accuracy of the numerical calculation. 
We arrange 128 replicas having different temperatures ranging from $T_{\rm c}(0)/4$ to $2T_{\rm c}(0)$, where $T_{\rm c}(0)$ is the transition temperature of the 3d model without field calculated by the 3dMC method. We take $10^5$--$10^8$ MC sweeps depending on the parameters to create the thermal equilibrium state, and take the same number of the MC samplings. 
The statistical uncertainties in our simulations are determined by standard deviations of 24 independent simulations with different initial conditions.

In the 3dMC calculation under the external magnetic field, we suffer from extremely slow convergency and many metastable states characterized by the quantized winding number under the periodic boundary condition. 
The transition between the states with different soliton densities is hardly realized in our algorithm and consequently the results are dependent on the initial conditions. 
Hence in this paper we apply the 3dMC method only to the case without external field, where the results such as the specific heat are insensitive to the choice of initial states. 
In the following calculations, we choose random spin configurations as the initial conditions.

\section{Results\label{section:results}}
\subsection{3dMF results}

In this section, we investigate the qualitative properties of the 3d system using the 3dMF method.
Here we take 4,000 sites along the $z$-axis. 
We impose the periodic boundary condition along the $z$-axis; for comparison, we also impose the open boundary condition along the axis (not shown here) and remark the superiority of the periodic boundary condition in Appendix.

\begin{figure}[tb]
	\begin{center}
		\includegraphics[width=9cm,clip]{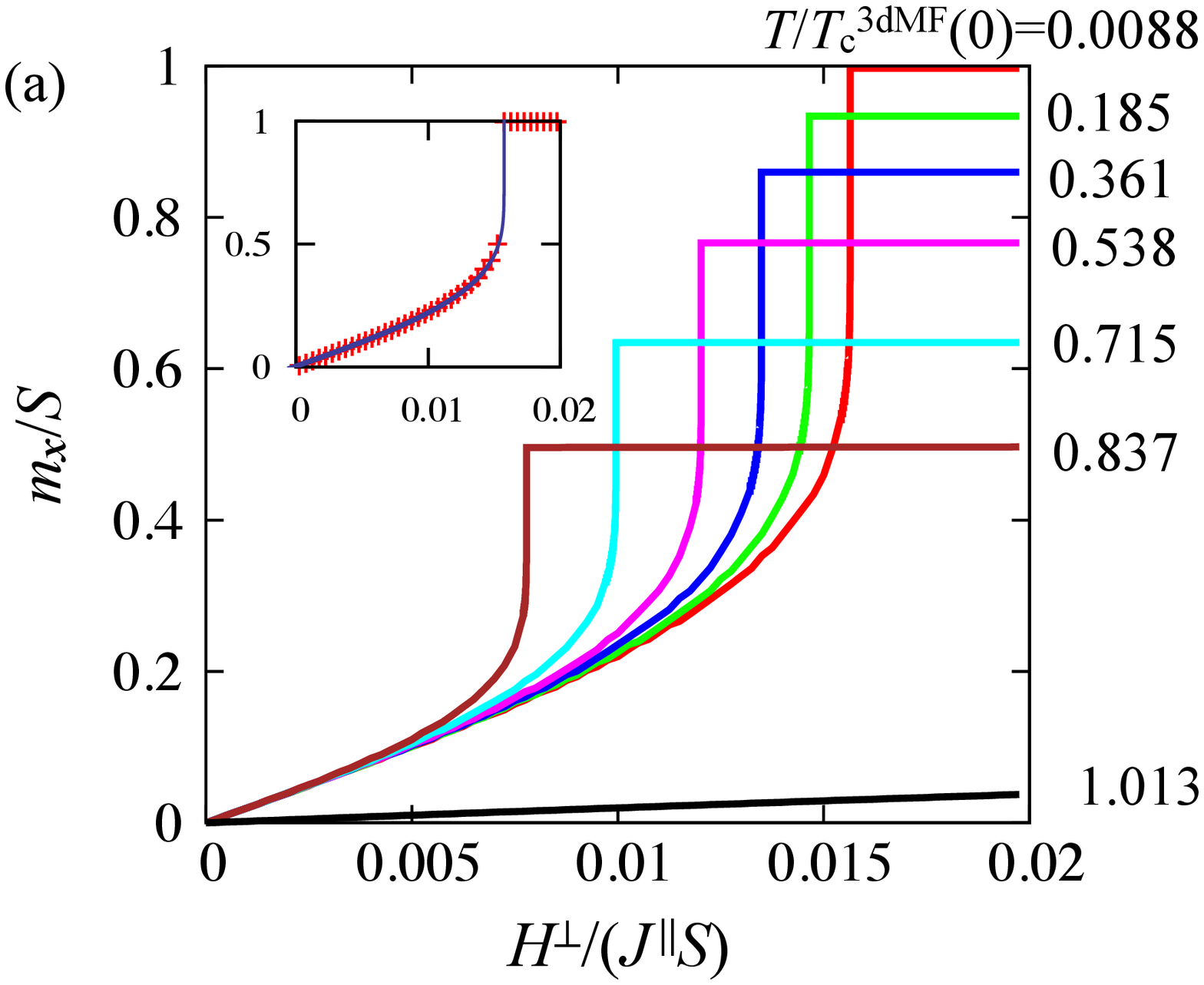} \label{MF_MHcurve}
		\includegraphics[width=9cm,clip]{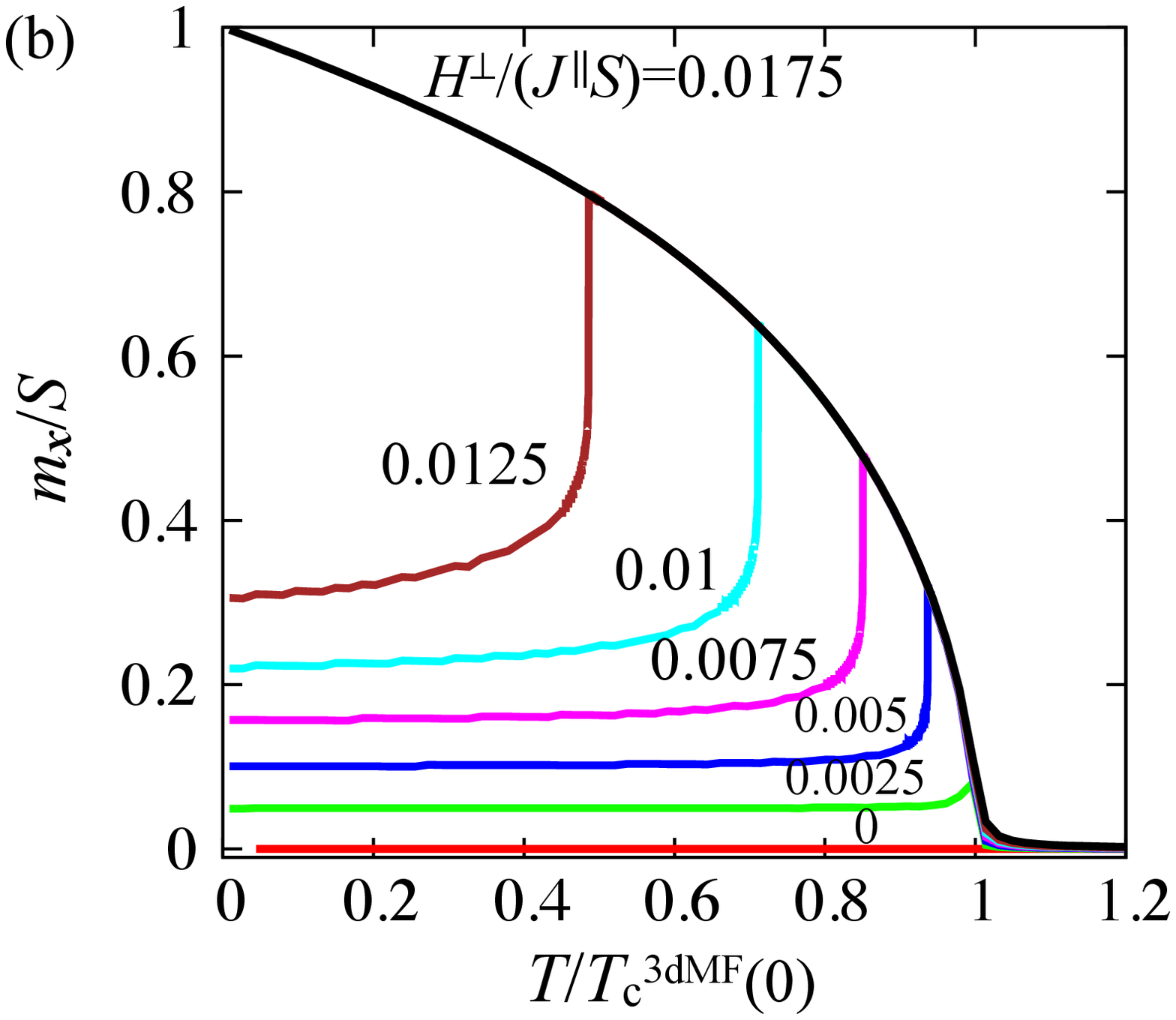} \label{MF_MTcurve}
	\end{center}
	\caption{Magnetization curves with $J^\bot/J^\parallel=8$, $D/J^\parallel = 0.16$ calculated by the 3dMF method. Here, $T_{\rm c}^{\rm 3dMF}(0)/(J^\parallel S^2)=11.34$. (a) Magnetization vs. applied field for several values of temperature. The inset shows the magnetization curves calculated by using the 3dMF method (red plots) and 1d analysis (blue solid line). (b) Magnetization vs. temperature for several values of external field.}
	\label{MF_MFMT}
\end{figure}

Figure~\ref{MF_MFMT}(a) shows the magnetization curves for several values of temperature. The horizontal and vertical axes are, respectively, the external field and the $x$-component of the magnetization ${\bm m}/S = \sum_{\bm i} {\bm M}_{\bm i} /(NS)$. Note that $m_y=m_z=0$ because the external field is applied along the $x$-axis. We choose $D/J^\parallel = 0.16$ and $J^\bot/J^\parallel = 8$, which are relevant to \Cr, as will be discussed in Sec.~\ref{section:discussion}. The inset shows that the magnetization curve obtained by the 3dMF method at $T=0.0088T_{\rm c}^{\rm 3dMF}(0)$ coincides with that of the 1d analysis based on the chiral sine-Gordon model~\cite{Kishine_S_T}. At $T>T_{\rm c}^{\rm 3dMF}(0)$, the field dependence of the magnetization is approximately linear and the gradients of the magnetization curves with small field fit to the Curie-Weiss law: $\partial m_x/\partial H^\bot|_{H^\bot=0} = S^2/[3(T-T_{\rm c}^{\rm 3dMF}(0))]$ for 3d Heisenberg model. We confirm the validity of our result through these observations. 

Below $T_{\rm c}^{\rm 3dMF}(0)$, the $m_x(H^\bot)$ curves exhibit characteristic convex downward behavior, which is observed in experiments~\cite{Miyadai,Ghimire,Chapman,Miyazaki, Ohara}. An inspection of the MF solution reveals that the CSL is formed below the $T$-dependent critical field $H_{\rm c}^\bot(T)$, above which the magnetization becomes uniform. 
Those spatially uniform states under high fields at low temperatures have been called a forced ferromagnetic region, which is continuously connected with paramagnetic region under low fields at high temperatures. Although the magnetization curves at high temperatures do not seem to be saturated at high field, we confirm that magnetization in the forced ferromagnetic region approaches $m_x/S=1$ when Zeeman energy is much larger than the thermal energy ($k_{\rm B}T$) [not depicted here]. 

Figure~\ref{MF_MFMT}(b) shows the temperature dependence of the magnetization for several values of magnetic field. 
We can see characteristic cusp structures near the transition temperature; a similar cusp has been observed in experiments~\cite{Miyadai,Ghimire,Ohara,Dyadkin} and has been obtained in a phenomenological theory~\cite{Kishine_MTcusp}. 
Small step-like structures at low temperatures are due to the finite-size effect along the helical axis.

\subsection{Spin-correlation effect on $T_{\rm c}(H^\bot=0)$}

\begin{figure}[tb]
	\begin{center}
		\includegraphics[width=9cm,clip]{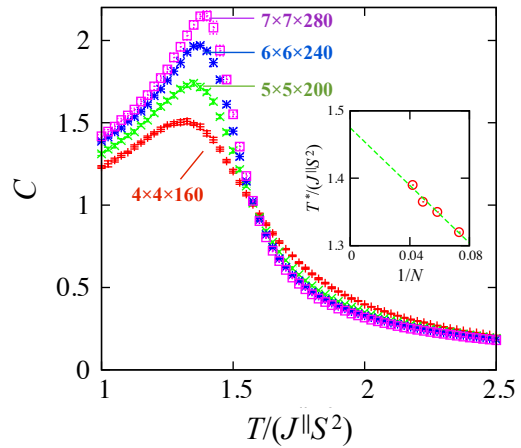}
	\end{center}
	\caption{Temperature dependence of the specific heat without external field for some system sizes: $(N_x, N_y, N_z) = (4,4,160), (5,5,200), (6,6,240),$ and $(7,7,280)$. Parameters are $J^\bot/J^\parallel=1$ and $D/J^\parallel=0.16$. The error bar of each point is much smaller than the symbol size. The inset shows the inverse of the system size $1/N$ vs. peak points $T^*$ of the specific heat.  The green dot line is a linear fitting function. }
	\label{CT_curve}
\end{figure}

We discuss quantitatively the spin-correlation effect on the transition temperature at zero field. First, we investigate the results obtained by the 3dMC method which fully includes spin correlations. In the simulation, we impose the periodic boundary condition. 
Anomalies at the phase transition can be seen in the specific heat $C(T)$. Figure~\ref{CT_curve} shows the temperature dependence of the specific heat for several system sizes. We choose $D/J^\parallel = 0.16$, which is relevant to \Cr\ as will be discussed in Sec.~\ref{section:discussion}. 
The larger the system size is, the sharper the peak of the specific heat becomes. These peak points $T^*$ slightly shift to higher temperature when we take the larger size of the system. 
In the inset of Fig.~\ref{CT_curve}, we plot $1/N$ $(N^3=N_x N_y N_z)$ dependence of the peak position in $C(T)$. By extrapolating the values to $N\rightarrow \infty$, we obtain the transition temperature at zero field in the thermodynamic limit. 
For example, in the case with $J^\bot/J^\parallel=1$, we obtain $T_{\rm c}(0)/(J^\parallel S^2) =1.47$, which is slightly higher than that of the classical Heisenberg model: $T_{\rm c}^{\rm Heisenberg}(0)/(JS^2)=1.44$~\cite{Chen}.

\begin{figure}[tb]
	\begin{center}
		\includegraphics[width=8.5cm,clip]{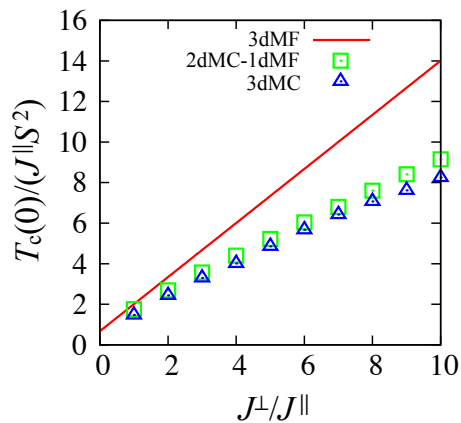}
	\end{center}
	\caption{Exchange interaction ratio vs. transition temperature without external field. The red line is the 3dMF solution (Eq.~(\ref{MF_Tc})), the green squares are the results calculated by the 2dMC-1dMF method, and the blue triangles are the 3dMC results. We set $D/J^\parallel=0.16$. The error bar of each point is much smaller than the symbol size.}
	\label{Tc_J}
\end{figure}

Let us compare the results obtained by the 3dMF method with those by the 3dMC study. The exchange interaction dependence of the transition temperature without external field is shown in Fig.~\ref{Tc_J}. As increasing $J^\bot/J^\parallel$, the 3dMC results (blue triangle plots) deviate from the 3dMF solution (red solid line). Because of a lack of the spin-correlation effect, the transition temperature is overestimated in the MF method. 

To improve the MF results quantitatively, we use 2dMC-1dMF method to calculate $T_{\rm c}(0)$. We take several system sizes $(N_x, N_y) = (20, 20), (50,50), (100,100)$ for the $xy$-plane, and confirmed that the system-size dependence is smaller than statistical errors.
The green square symbols in Fig.~\ref{Tc_J} are the results calculated by the 2dMC-1dMF method based on Eq.~\eqref{eq:2d_MF}. 
The evaluation of $T_{\rm c}(0)$ is highly improved from the 3dMF description by taking account of the spin-correlation within 2d layers.

\subsection{Spin-correlation effect with external field}

Here we discuss the spin-correlation effect on the system with external field by comparing the results of 3dMF method with those of 2dMC-1dMF method. Here we take 2,000 sites along the $z$-axis for these two methods, and take $(N_x, N_y) = (20, 20)$ and $(50,50)$ for the $xy$-plane in the 2dMC-1dMF calculation. 
The system-size dependence of the 2d layer is not essential, because it is the number of layers that mainly determines the numerical accuracy.

\begin{figure}[tb]
	\begin{center}
		\includegraphics[width=8cm,clip]{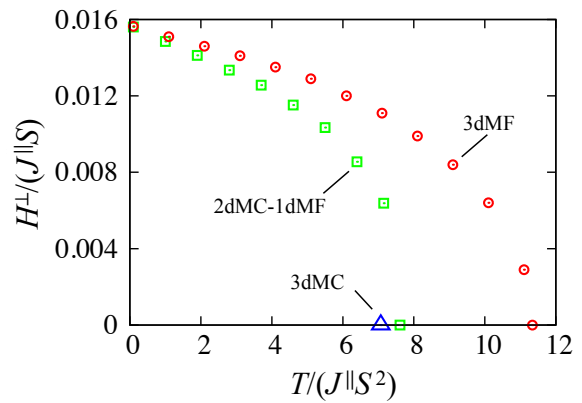} 
	\end{center}
	\caption{Phase diagrams calculated by the 3dMF and 2dMC-1dMF methods with $D/J^\parallel = 0.16$, $J^\bot/J^\parallel =8$. The error bar of each point is much smaller than the symbol size. The blue triangle denotes 3dMC result for the transition temperature at zero field. }
\label{phase_MFvs1dMF}
\end{figure}

\begin{figure}[tb]
	\begin{center}
		\includegraphics[width=9cm,clip]{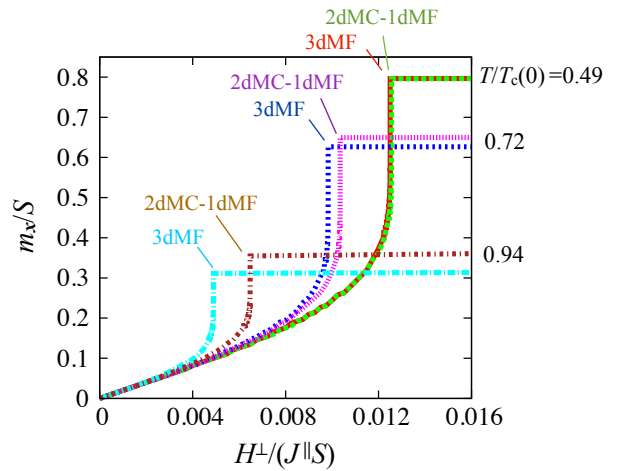} 
	\end{center}
	\caption{Magnetization vs. applied field with several values of temperature $T/T_{\rm c}(0)$ calculated by the 3dMF and 2dMC-1dMF methods. Here we choose $D/J^\parallel = 0.16$, $J^\bot/J^\parallel =8$ where the transition temperature at zero field is $T_{\rm c}^{\rm 3dMF}(0) / (J^\parallel S^2) = 11.34$ for the 3dMF method and $T_{\rm c}^{\rm 2dMC\mathchar`-1d MF}(0) / (J^\parallel S^2) = 7.61$ for the 2dMC-1dMF method.}
\label{MHcurve_MFvs1dMF}
\end{figure}

We plot the phase diagrams calculated by these two methods in Fig.~\ref{phase_MFvs1dMF}. The critical field at zero temperature $H_{\rm c}^\bot(0)$ has the same value because the spin-correlation effect vanishes. While the transition temperature obtained by the 3dMF method is overestimated also at finite external field, the qualitative behaviors of the phase diagrams are similar for these two methods.
Figure~\ref{MHcurve_MFvs1dMF} shows the magnetization curves for different temperatures, whose values are scaled by $T_{\rm c}(0)$, calculated by the 3dMF and 2dMC-1dMF methods. At sufficiently low temperatures, these two results agree with each other. At higher temperatures such as $T/T_{\rm c}(0)=0.72$ and $0.94$, the results of two methods give different critical points, which indicates that the phase boundaries do not match with each other even if we scale the temperature by $T_{\rm c}(0)$.

%

\section{Discussion\label{section:discussion}}

In this section, we discuss implications for \Cr\ from our studies. We have already known some characteristic parameters of \Cr\ from the experiment~\cite{Miyadai}, such as the lattice constant $a_0 = 1.2~{\rm nm}$, the spatial period of the helix $L=48~{\rm nm}$, the transition temperature at zero field $T_{\rm c}(0)= 127~{\rm K}$, and the critical field at low temperature $H^\bot_{\rm c}(0) \simeq 0.14~{\rm T}$ ($=0.19~{\rm K}$ with $g$-factor $g=2$). By comparing our numerical results with these experimental data, we can estimate the interaction parameters $J^\bot$, $J^\parallel$ and $D$ in \Cr.

First, let us recall an estimation based on the 1d chiral sine-Gordon model~\cite{Kishine_Hc_perp}. There is a relation between the period $L$ and the DM interaction: $L/a_0 = 2\pi / \arctan(D/J^\parallel)$. Experimentally observed value~\cite{Miyadai} yields $D/J^\parallel=0.16$. 
Using the analytical expression of the critical field at zero temperature written by $H_{\rm c}^\bot(0)/(J^\parallel S) = \left(\pi\alpha/4 \right)^2$~\cite{Kishine_Hc_perp}, we derive $J^\parallel S^2=18~{\rm K}$, and $DS^2=2.9~{\rm K}$. We have no information on $J^{\perp}$ in the 1d model.

Next, we estimate the set of parameters including $J^\perp$ with our model.
We have observed no temperature dependence of the period $L$ below $T_{\rm c}(0)$ by the 3dMC method, and it well agrees with the chiral sine-Gordon analysis. 
Hence we also set $D/J^\parallel = 0.16$ for 3d systems. 
From the 3dMF results, which are valid at sufficiently low temperatures, we estimate $H_{\rm c}^\bot(0)/(J^\parallel S)= 0.0157$ for $D/J^\parallel=0.16$. We then obtain $J^\parallel S^2 = 18~{\rm K}$ and $D S^2 = 2.9~{\rm K}$, which are the same values as the estimation using the chiral sine-Gordon model. The intra-layer exchange interaction $J^{\perp}$ is related to the transition temperature.
From Fig.~\ref{Tc_J} we obtain $J^\bot S^2 = 86~{\rm K}$ for the 3dMF description, and $J^\bot S^2 = 1.3 \times 10^2~{\rm K}$ for the 2dMC-1dMF approach. The 3dMC results also provide us an estimation: $J^\bot S^2 = 1.4 \times 10^2~{\rm K}$, which is the most realistic value among these three methods.
These values are derived from comparison between experiments and numerical results on our simple model based on the Hamiltonian [Eq.~\eqref{eq:hamiltonian}] on the cubic lattice; the realistic situation such as the lattice structure is, however, different from the present setup in various ways.

The result for $J^{\parallel} \gg D$ gives the very gradual change of the spin texture, and the condition $J^{\bot} \gg J^{\parallel}$ quantitatively justifies the quasi-2d picture:
the Cr atoms in this crystal are strongly coupled together within the $ab$-plane, and are weakly correlated between these planes. 
The strong spin-coupling in the $ab$-plane originates from the 2d layer structure of NbS$_2$. The high transition temperature $T_{\rm c}(0) = 127~{\rm K}$ of \Cr\ stems from this strong intra-layer exchange interaction.

\section{Conclusion\label{section:conclusion}}
We have analyzed the 3d chiral helimagnets using the MF method to investigate the physical properties at finite temperatures. The simple Hamiltonian with the Heisenberg exchange interaction, the DM interaction, and the magnetic field can qualitatively describe the magnetic properties of chiral helimagnets at zero and finite temperatures. 

For the quantitative investigation, it is important to include the spin-correlation effect. We have calculated physical quantities for the system at zero field using the 3dMC method, and have shown that the transition temperature estimated by using the 3dMF method deviates from the 3dMC results as a consequence of  quasi-two-dimensionality. Using the hybrid method called 2dMC-1dMF method, we have taken account of the spin-correlation effect in the plane perpendicular to the helical axis. The estimated transition temperatures are much improved from the simple MF theory.
This result implies that the 2dMC-1dMF method enables us to quantitatively investigate the thermodynamic properties under external field with less numerical effort than the 3dMC method. 

We have estimated the three interaction parameters in \Cr\ as $D=2.9$~K, $J^\parallel=18$~K and $J^\perp=1.4\times 10^2$~K based on our simple model. 
The strong intra-layer exchange interaction is closely related to the high $T_{\rm c}$ in \Cr. 

Critical phenomena near the phase transition between CSL and paramagnetic states are an open issue. \\

{\it Note added}. After completing the first version of our manuscript [arXiv1512.00235v2], we come to know an experimental result suggesting a tri-critical point, which is a boundary between the first- and second-order phase transition, at about 40\% of $H^\bot_{\rm c}(0)$~\cite{Tsuruta}. 
The change of the order/type of the phase transition along the phase boundary in the magnetic field was also discussed recently by two groups~\cite{Victor, Nishikawa}. We briefly remark on the order of the transition in our analysis. Around the phase boundary at $H^\bot /(J^\parallel S)=0.0006,0.0007,0.0008,0.0009, 0.001$, we find double minimum structures in the free energy as a function of the winding number (see Appendix). This observation implies the first-order phase transition. 
The double minimum structures are observed only in narrow temperature region; e.g. at $H^\bot /(J^\parallel S)=0.0006$ where $T_{\rm c}(H^\bot)/(J^\parallel S^2)=11.3312328125 \pm 0.0000003125$, the double minimum structures are observed in $11.3309775<T/(J^\parallel S^2)<11.3320475$.

\begin{acknowledgments}
We thank K.~Hukushima for critical reading of the manuscript and T.~Takahashi for his advice on Monte Carlo algorithm. We also thank Y.~Togawa, A.~Ovchinnikov and Y.~Nishikawa  for their informative discussions. A part of computation in this work has been done using the facilities of the Supercomputer Center, the Institute for Solid State Physics, the University of Tokyo. This work was supported by the Center for Chiral Science in Hiroshima University.
\end{acknowledgments}
\appendix
\section{The order of the transition}

%
\begin{figure}[tb]
	\begin{center}
		\includegraphics[width=7cm,clip]{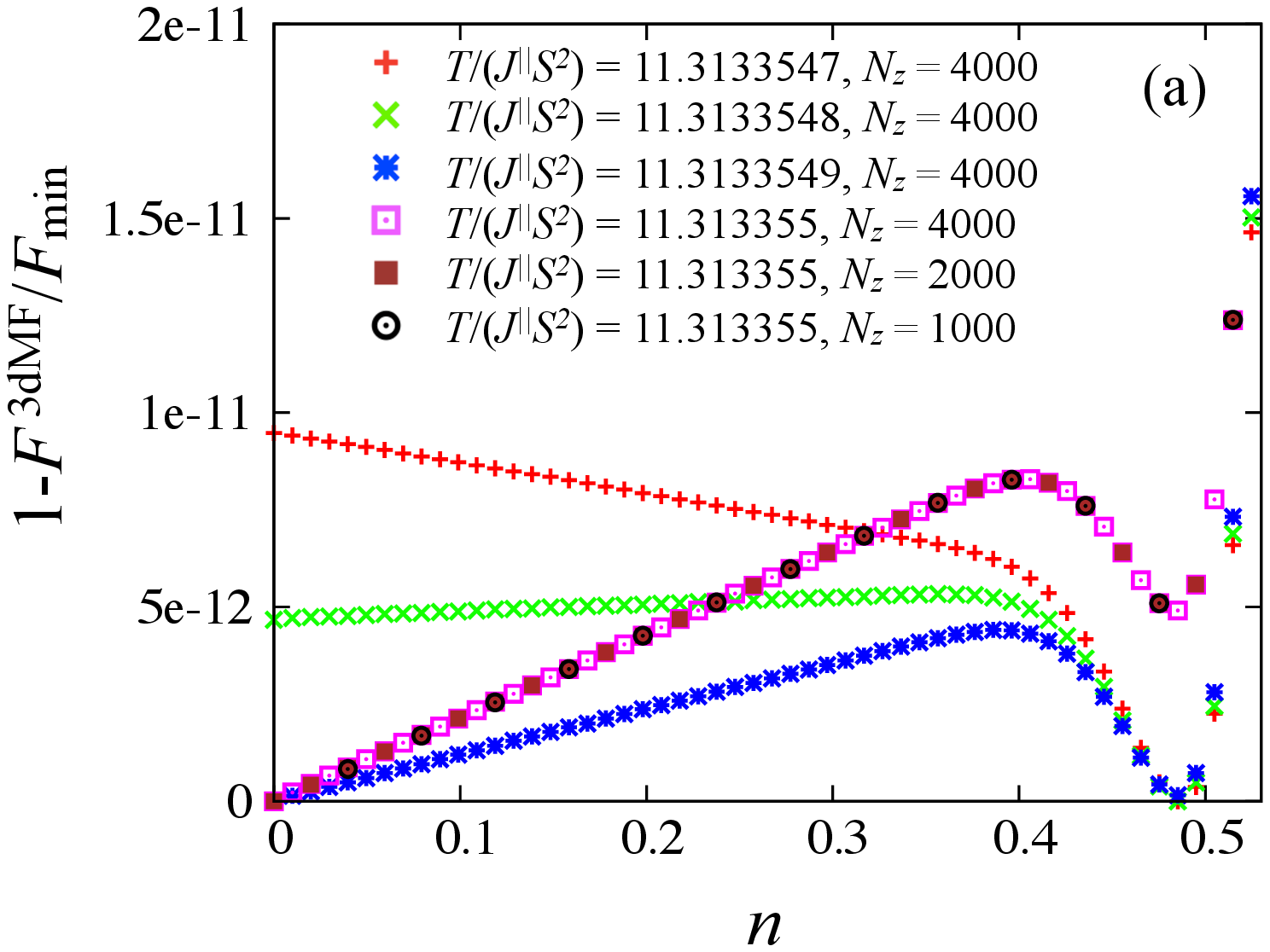}
		\includegraphics[width=7cm,clip]{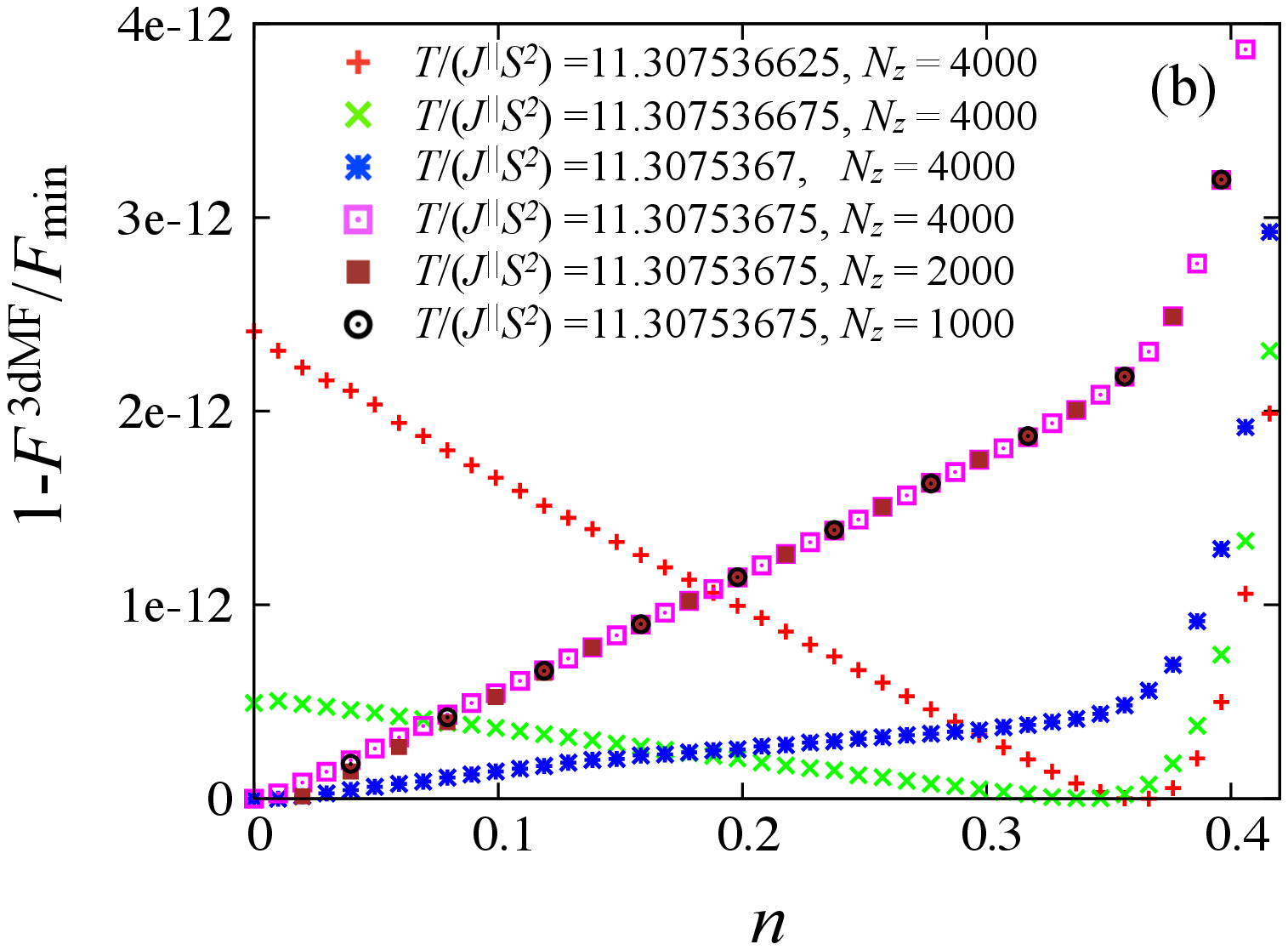}
	\end{center}
	\caption{Winding-number ratio dependence of the free energy at $H^\bot/(J^\parallel S ) = 0.001$ (a), and $H^\bot/(J^\parallel S ) = 0.0011$ (b) for several values of temperature near the transition point. Here, $F_{\rm min}$ is the minimum free-energy for given temperature and magnetic field. Parameters are set $D/J^\parallel=0.16$ and $J^\bot/J^\parallel=8$.}
\label{fig:appendix}
\end{figure}

Figure~\ref{fig:appendix} (a) shows a result of the free energy that implies a first-order transition, obtained by the 3dMF method. Using the winding number at zero field $w_0$, we introduce the winding-number ratio $n$ defined by $ n = w_{\rm f}/w_0$. 
Here $w_{\rm f}= \sum_{i_\parallel} \Delta \theta_{i_\parallel}/(2\pi)$ is the winding number which is integer under the periodic boundary condition, and $\Delta \theta_{i_\parallel}$ is the angle between the spins located at the nearest neighbor layers written as 
\begin{align}
	\Delta\theta_{i_\parallel} = \arcsin\left[
		\frac{M^x_{i_{\parallel}+1}M^y_{i_\parallel}-M^y_{i_{\parallel}+1}M^x_{i_\parallel}}
			{	\sqrt{(M^{x}_{i_{\parallel}})^2 + (M^{y}_{i_{\parallel}})^2}
				\sqrt{(M^{x}_{i_{\parallel}+1})^2 + (M^{y}_{i_{\parallel}+1})^2}
			}
	\right].
\end{align}
We see double minimum structures, so that the free-energy minimum jumps from $n\simeq 0.5$ to $n=0$ with increasing temperature. 

Figure~\ref{fig:appendix} (b) shows $n$ dependence of the free energy at $H^\bot/(J^\parallel S ) = 0.0011$ for several values of temperature near the transition point. 
The free energy minimum gradually shifts to smaller $n$ with increasing temperature, and the free-energy curve becomes broad against $n$ at around $T_c(H^\bot)$. After that, the minimum always exists at $n =0$. We do not observe the double minimum at $H^\bot=0$ and $0.011<H^\bot /(J^\parallel S) <H_{\rm c}^\bot(0) /(J^\parallel S)$ in the range of $|T - T_c(H^\bot)| /(J^\parallel S^2) > 2.5 \times 10^{-8}$. 

Under the periodic boundary condition, the size dependence of the free energy is so small that the results for $N_z=4000, 2000,1000$ collapse onto a single curve as shown in Fig~\ref{fig:appendix}.
Under the open boundary condition, on the other hand, we do not observe the double minimum structures of the free energy. It is because the system-size dependence of the free energy remains even in larger systems with $N_z=8000, 4000,1000$; the free-energy profile under the open boundary condition approaches that under the periodic boundary condition with increasing the system size. It reveals that the physical property in the thermodynamic limit is much more accessible under the periodic boundary condition. 

\bibliography{apssamp}
\bibliographystyle{prsty}

\end{document}